\documentclass[fleqn,10pt]{wlscirep}
\usepackage[utf8]{inputenc}
\usepackage[T1]{fontenc}
\title{Coexisting Ordered States, Local Equilibrium-like Domains, and Broken Ergodicity in a Non-turbulent Rayleigh-B{\'e}nard Convection at Steady-state}

\author[*]{Atanu Chatterjee}
\author[+]{Yash Yadati}
\author[+]{Nicholas Mears}
\author[ ]{Germano Iannacchione}
\affil[ ]{Department of Physics and the Order-Disorder Phenomena Lab, Worcester Polytechnic Institute, MA, USA, 01605}

\affil[*]{achatterjee3@wpi.edu}

\affil[+]{these authors contributed equally to this work}

\begin{abstract}
A challenge in fundamental physics and especially in thermodynamics is to understand emergent order in far-from-equilibrium systems. While at equilibrium, temperature plays the role of a key thermodynamic variable whose uniformity in space and time defines the equilibrium state the system is in, this is not the case in a far-from-equilibrium driven system. When energy flows through a finite system at steady-state, temperature takes on a time-independent but spatially varying character. In this study, the convection patterns of a Rayleigh-B{\'e}nard fluid cell at steady-state is used as a prototype system where the temperature profile and fluctuations are measured spatio-temporally. The thermal data is obtained by performing high-resolution real-time infrared calorimetry on the convection system as it is first driven out-of-equilibrium when the power is applied, achieves steady-state, and then as it gradually relaxes back to room temperature equilibrium when the power is removed. Our study provides new experimental data on the non-trivial nature of thermal fluctuations when stable complex convective structures emerge. The thermal analysis of these convective cells at steady-state further yield local equilibrium-like statistics. In conclusion, these results correlate the spatial ordering of the convective cells with the evolution of the system's temperature manifold.
\end{abstract}
\begin{document}

\flushbottom
\maketitle
%
%
\thispagestyle{empty}


\section*{Introduction}

\begin{figure}[t]
    \centering
    \includegraphics[scale=0.8]{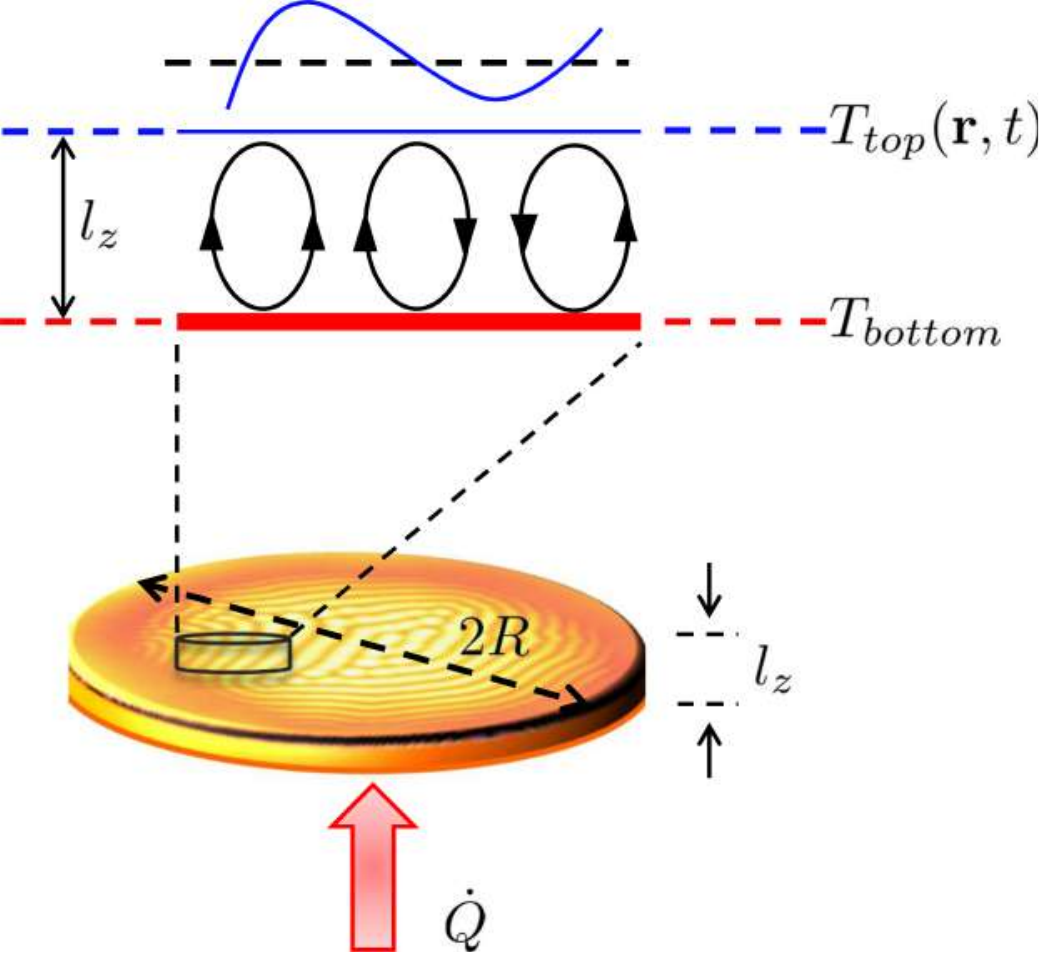}
    \caption{Cartoon illustrates the experimental configuration of the current study. The Rayleigh-B{\'e}nard system at steady-state is set up by heating a thin film of viscous liquid from the bottom ($\dot{Q}$). The temperature difference between $T_{bottom}$ and $T_{top}$ gives rise to convection rolls. While at steady-state, $T_{bottom}$ is constant, real-time thermal imaging of the top layer is performed to extract the spatial and temporal distribution of $T_{top}$. The line cut of the thermal profile $T_{top}(\mathbf{r},t)$ is also shown. As the goal was to have convection cells over as wide as an area possible for the thermal imaging to yield significant temperature statistics, a large diameter to thickness ratio of the apparatus ($2R / l_z$ $\simeq$ $225$ $mm$ / $5$ $mm$ $\sim$ $45$) yielded a stable convection cell pattern $\gtrsim 150$ $mm$ in diameter and stable for as long as the power was applied.}
    \label{fig10}
\end{figure}

From swarming in biological organisms to crack propagation in materials, from phase-transitions to glass transitions, from molecular processes at nanoscale that form the basis of life to the ever-changing climate on this planet, it is no coincidence that everything around us operate under conditions that are far-from-equilibrium. These systems, that have been driven out-of-equilibrium exhibit an incredibly wide variety of patterns that emerge spontaneously through local interactions. By virtue of being driven out-of-equilibrium, these systems are typically nonlinear, thermodynamically open, often non-ergodic and disordered while exhibiting spontaneous emergent order at the same time. The study of such systems, therefore becomes an extremely challenging affair~\cite{cross1993pattern, kadanoff2001turbulent, jaeger2010far, egolf2002far, chatterjee2016energy, chatterjee2017aging, lucia2013stationary}. In order to gain insights about these out-of-equilibrium systems some model systems that are actively studied include clustering of bacterial colonies and self-assembly in actomyosin motility assays, phase ordering in liquid crystals, synchronization of Kuramoto oscillators, oscillatory behaviors in reaction-diffusion systems such as the Belousov-Zhabotinsky reaction, or turbulence and pattern formation in thermal-convective systems like the Rayleigh-B{\'e}nard convection~\cite{cross1993pattern, huber2018emergence, kuramoto1987statistical, zhang1993deterministic, behringer1985rayleigh, koschmieder1993benard, bejan2013convection}.

In this paper, we focus on the Rayleigh-B{\'e}nard convection as a prototype for a far-from-equilibrium system that exhibits emergent order. It should be noted that our study of the Rayleigh-B{\'e}nard convection is motivated \emph{solely} from a thermodynamic point of view and not from a fluid mechanics perspective. To elaborate, we focus on broad questions such as can multiple local equilibrium states coexist in an otherwise far-from-equilibrium system, or how the statistical mechanics of a far-from-equilibrium system differs from that of a system at equilibrium? What are the limitations of the local equilibrium hypothesis, or under what conditions do thermal gradients in a system dominate and allow for the spontaneous emergence of ordered structures~\cite{cross1993pattern, kolmogorov1941local, vilar2001thermodynamics, chatterjee2018many, chatterjee2016thermodynamics, georgiev2016road, Verma2018, kumar2018physics,verma2019asymmetric}? While we experimentally explore the far-from-equilibrium behavior of temperature, these results sheds light on the fundamental questions mentioned above. These questions, answers to which are yet unknown or inconclusive, are important for the broad scientific community, but are also of significant general interest.

The Rayleigh-B{\'e}nard convection, due to its conceptual richness and an easy experimental methodology, remains one of the most actively and extensively studied physical system. The dynamics of a Rayleigh-B{\'e}nard convection system joins fundamental ideas from both thermodynamics and fluid mechanics. When a thin film of liquid is heated, the competing forces of viscosity and buoyancy give rise to convective instabilities. This convective instability creates a spatio-temporal non-uniform thermal distribution on the surface of the fluid film. The advantage of this system lies in its simplicity, wherein a dimensionless quantity, the Rayleigh number $(Ra)$, determines the onset of convective cell patterns~\cite{cross1993pattern, koschmieder1993benard, rayleigh1916lix}. The critical Rayleigh number of $1708$ marks the onset of convection for a no-slip boundary condition was obtained by Jeffreys in 1929~\cite{cross1993pattern, koschmieder1993benard, rayleigh1916lix}. Since then there has been a series of studies on the empirical relationships between the various dimensionless numbers (specially, Nusselt's number ($Nu$), Reynold's number ($Re$), Prandtl's number ($Pr$) and $Ra$) under conditions of laminar and turbulent flows~\cite{zhong2009prandtl, stevens2011prandtl}. Studies reporting the efficiency in convective heat transfer based upon geometry or on the role of plumes, to measuring thermal fluctuations under turbulent flow conditions, have played an important role in understanding convection cell formation~\cite{du1998enhanced, du2000turbulent, du2001temperature, chilla2004ultimate, grossmann2004fluctuations, schumacher2008lagrangian, chilla2012new}. The onset of convection cell patterns in relation to thermal and hydrodynamic boundary layer models is an active area of interest in the fluid mechanics community, especially in exploring turbulence. Turbulence, although quite ubiquitous in nature, still remains one of the many unsolved problems in physics today. Not only as a tabletop experiment, but also through numerical simulations, the Rayleig-B{\'e}nard convection cell system serves as a very convenient prototypical model that has provided insights into the physics and hydrodynamics of turbulence. Noteworthy among them are studies on the effects of rotation and magnetic fields on Rayleigh-B{\'e}nard convection cells, turbulent convection at very high Rayleigh numbers with cryogenic $He$ gas as the working fluid to probe velocity and thermal statistics, and measurements of the mean temperature and variance profile as a function of boundary layer thickness~\cite{chandrasekhar2013hydrodynamic, niemela2000turbulent, shishkina2010boundary, stevens2010radial, wang2016boundary}.

This paper focuses on the non-turbulent spatio-temporal aspects of a Rayleigh-B{\'e}nard system at steady-state from a statistical physics point of view. Figure~\ref{fig10} illustrates the experimental configuration of the current study. A Rayleigh-B{\'e}nard system at steady-state is set up by heating at constant power ($\dot{Q}$) a thin film of a viscous liquid (oil) from the bottom. After stability is achieved, the temperature at the bottom of the oil film ($T_{bottom}$), which is in direct contact with the copper pan, becomes time-independent. The temperature of the top layer of the oil film ($T_{top}$), however exhibits spatial variation. This calorimetric information is extracted by performing real-time thermal imaging, and is further quantified by analyzing the spatio-temporal distributions of the thermal fluctuations. We note again that this paper \emph{does not} aim to uncover new physics that is of interest to the fluid mechanics community, but it aims to provide interesting insights into the physics of far-from-equilibrium thermodynamics. Our analysis provides insights about the distribution of thermal states when far-from-equilibrium. In the context of the analysis, we also discuss ergodicity and symmetry-breaking, and the local equilibrium hypothesis while experimentally determining the temperature distribution of a system driven out-of-equilibrium~\cite{jaeger2010far, casas2003temperature, martyushev2006maximum, lieb1998guide, lucia2008probability}. Through this paper we hope to spur theoretical interest in the description of far-from-equilibrium steady-state systems.

\begin{figure}[t]
    \centering
    \includegraphics[scale=0.6]{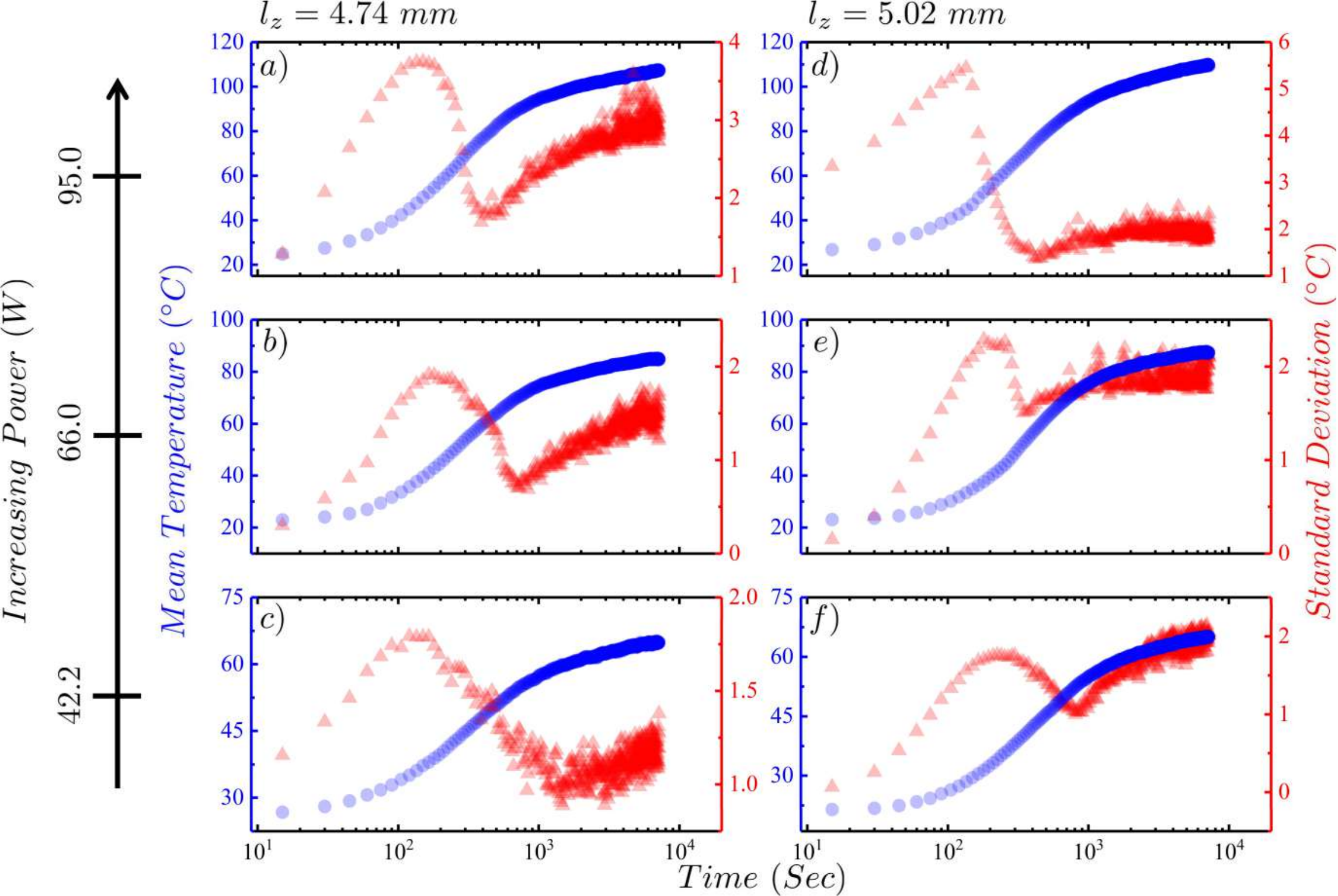}
    \caption{Figure shows on a semi-log scale the temperature mean and standard deviation as a function of time of the top of the silicone oil film as it responds to the applied heating power until steady-state is reached for various values of input power. The left axis corresponds to the temperature mean in degrees Celsius (solid blue circles) and the right axis corresponds to the standard deviation (solid red triangles). Plots $a$, $c$, $e$ show heating profiles for a film thickness of $l_z = 4.74$ $mm$, and plots $b$, $d$, $f$ for $l_z = 5.02$ $mm$. Note that the applied heating power in Watts are labeled by the far left $y-$axis.}
    \label{fig4}
\end{figure}

\section*{Results}

\begin{figure}[t]
    \centering
    \includegraphics[scale=0.6]{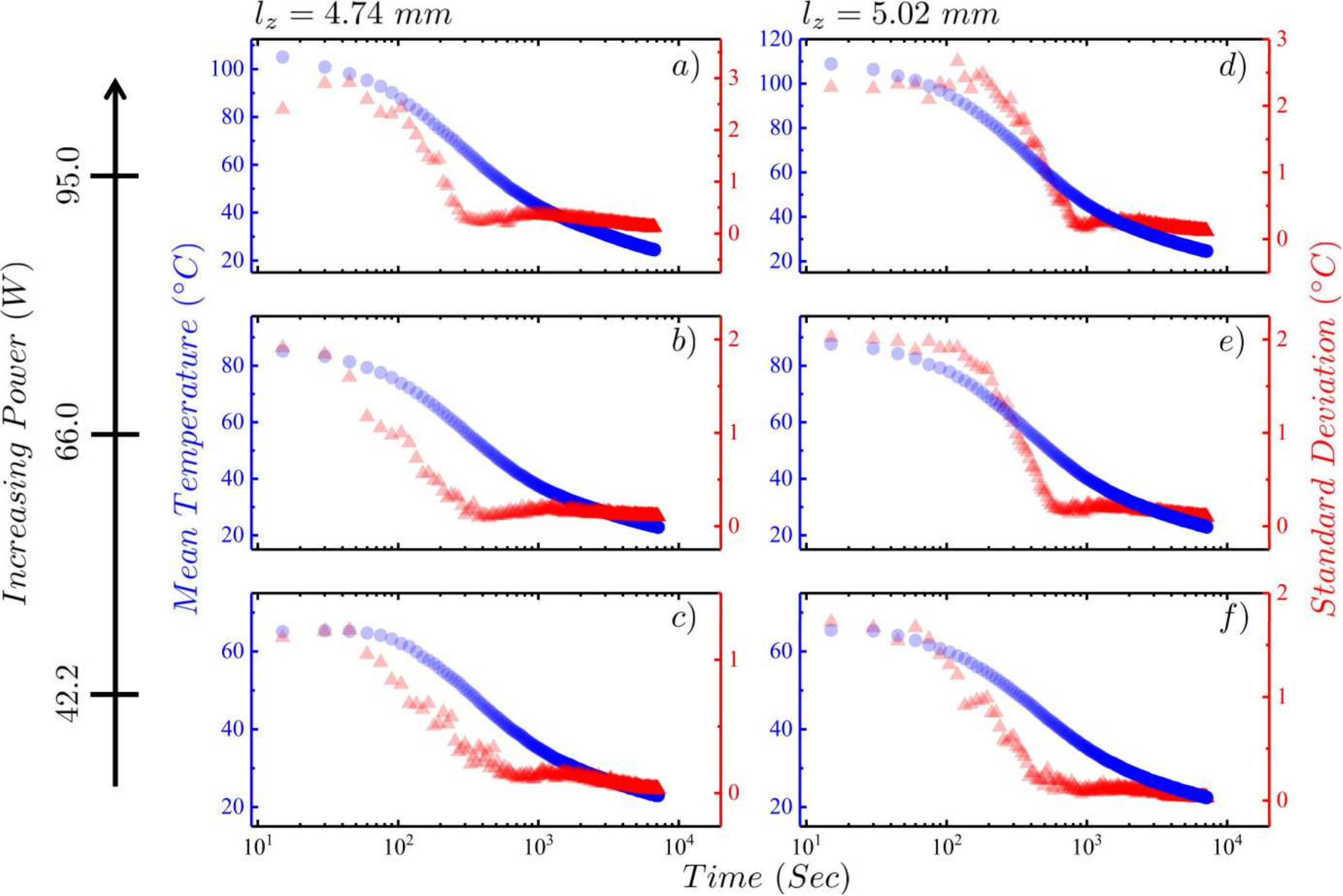}
    \caption{Figure shows on a semi-log scale the temperature mean and standard deviation as a function of time of the top surface of the silicone oil film as it relaxes to room temperature after the applied heating power is removed. The left axis corresponds to the temperature mean in degrees Celsius (solid blue circles) and the right axis corresponds to the standard deviation (solid red triangles). Plots $a$, $c$, $e$ show cooling profiles for the film thickness of $l_z = 4.74$ $mm$, and plots $b$, $d$, $f$ for $l_z = 5.02$ $mm$. Note that the applied heating power in Watts are labeled by the far left $y-$axis.}
    \label{fig5}
\end{figure}

\subsection*{Temporal Analysis}

In Figure~\ref{fig4} we plot the mean of the top temperature (left axis) and its standard deviation (right axis) as a function of time when the silicone oil sample is heated. The sample, initially at room temperature is driven out-of-equilibrium by the application of a constant heating power. Once the system reaches a steady-state, the heating power is switched off and system gradually relaxes back to room temperature. The top temperature mean and standard deviation as a function of time for the cooling process is plotted in Figure~\ref{fig5}. The mean temperature of an arbitrary region of interest on the image, $\langle T\rangle = \frac{1}{N}\sum_{i,j\in I}T_{ij}$ and the standard deviation, $\sigma_T = \sqrt{\frac{\sum_{i,j\in I}(T_{ij} - \langle T\rangle)^2}{N-1}}$ are calculated from the image matrix ($I_{ij}$). We observe from Figs~\ref{fig4} and \ref{fig5} that the mean temperature follows a typical heat-conduction trend for heating as the system achieves ostensibly a new high-temperature equilibrium as well as on cooling toward the original room temperature equilibrium state. The maximum temperature reached by each sample at steady-state increases as expected with increasing power based on the heat capacity for each film.

The plots of the temperature standard deviation as a function of time, however, show a markedly different trend during both heating and cooling processes as can be seen from Figures~\ref{fig4} and~\ref{fig5}, respectively. The standard deviation, a measure of the distribution width and is related to the temperature fluctuations in the system, generally increases with increasing temperature. Although an increasing trend in standard deviation as a function of time is observed on heating as expected since the temperature is increasing, this trend is broken at a point in time when the first hints of convection cells appear $\approx 200$ seconds, where $\sigma_T$ begins to decrease. This decrease in $\sigma_T$ continues as the convection cells grow until they reach their maximum extent over the film, which is not the entire film area due to the side heating produced by the Cu walls. Once the convection cell pattern has stabilized, $\sigma_T$ reaches a minimum at $\approx 900$ seconds after which it begins to increase again and only flattens as the mean temperature stabilizes. For cooling, after the heating power is removed, both the $\langle T\rangle$ and $\sigma_T$ begin to decrease with $\sigma_T$ decreasing more rapidly as time progresses until the last vestiges of any cell pattern disappears after which the decrease in $\sigma_T$ abruptly slows and flattens as $\langle T\rangle$ returns to room temperature. Over regions of the film where the temperature appears uniform, $\sigma_T$ is dominated by the spatial thermal fluctuations of the film but when convection cells are apparent $\sigma_T$ contains additional contributions due to thermal gradients across the film.

\begin{figure}[t]
    \centering
    \includegraphics[scale=0.6]{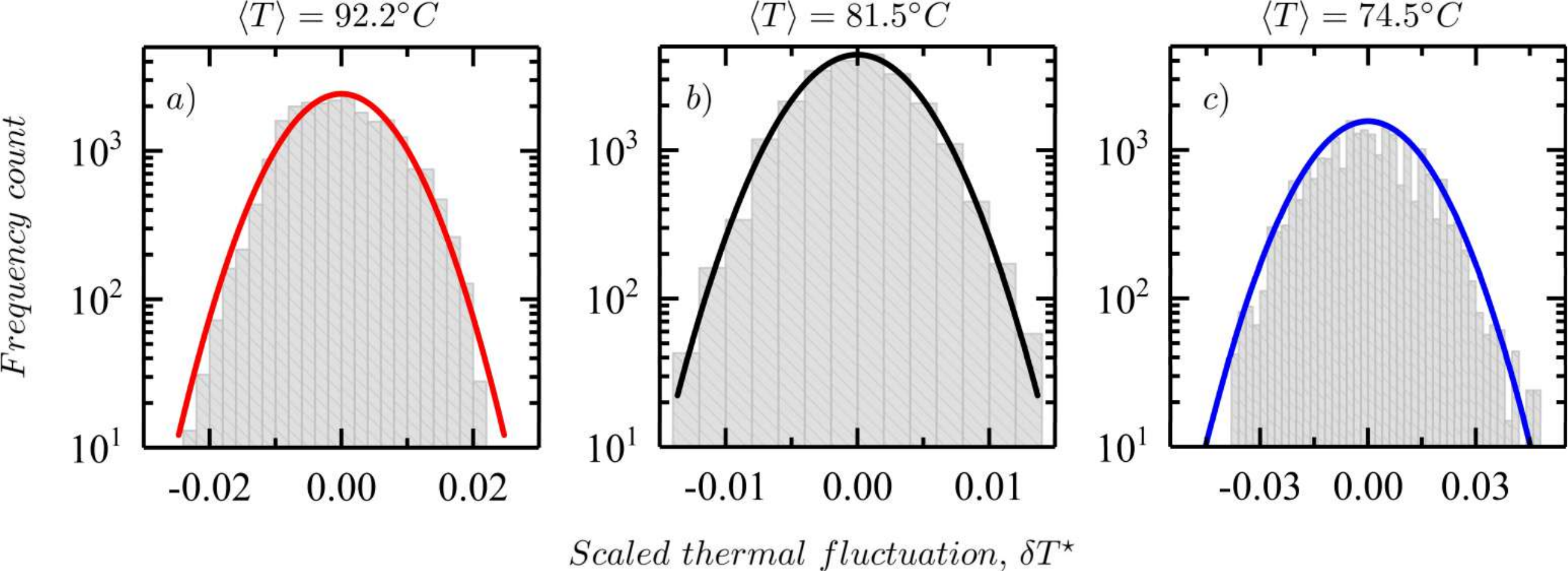}
    \caption{Figure shows the histograms for the scaled-thermal fluctuations averaged over time after the system has reached a steady-state on a semi-logarithmic scale. The panels a) denote the hot regions ($P_{hot}$), b) the entire region ($P$), and c) the cold regions ($P_{cold}$). The mean temperature, $\langle T\rangle$ (in $^\circ C$) of the various regions of interest are also denoted. The histograms are fitted with normal distribution functions all centered at zero.}
    \label{fig6}
\end{figure}

\begin{figure}[hb!]
    \centering
    \includegraphics[scale=0.6]{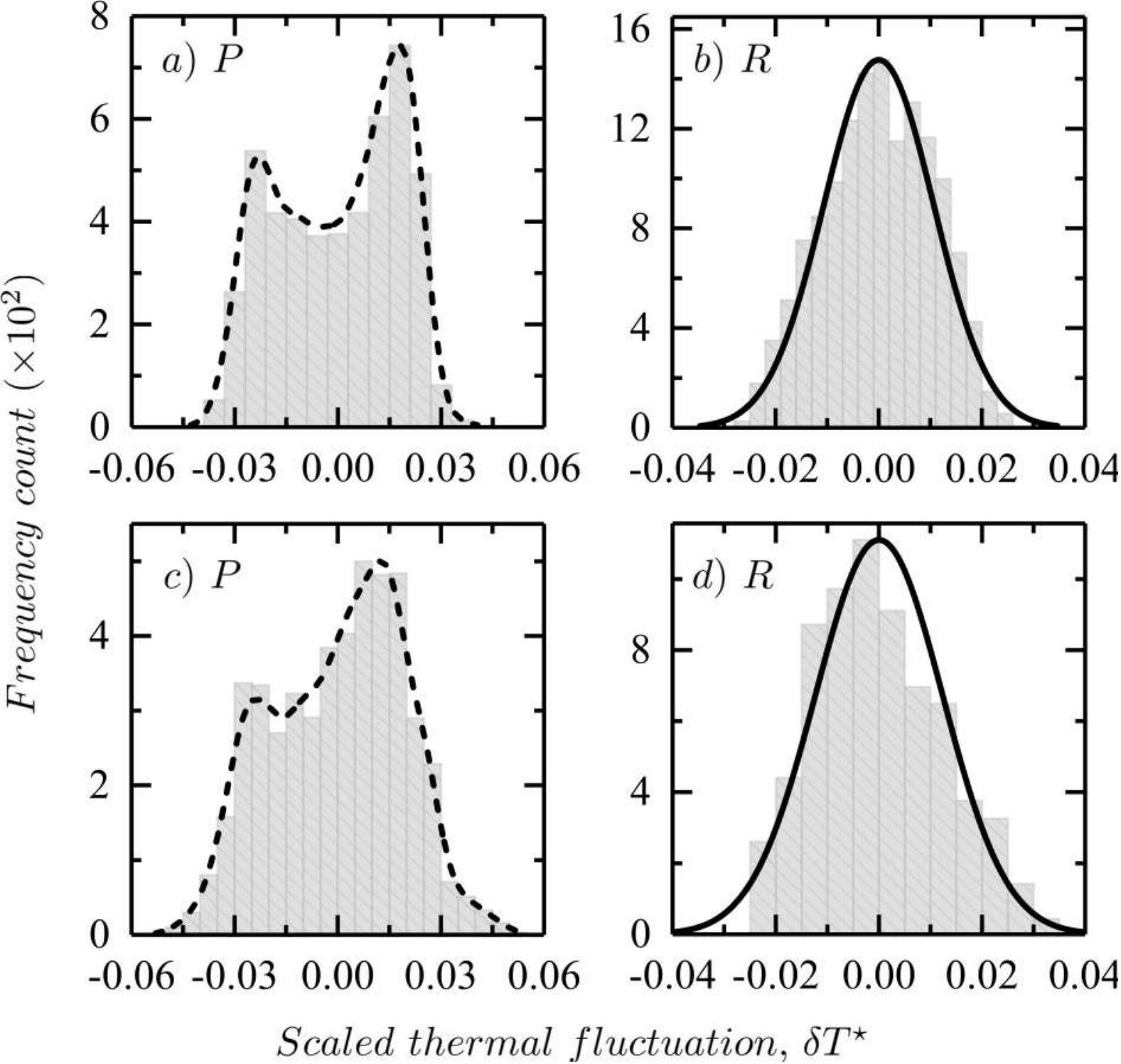}
    \caption{Figure shows the histograms for the scaled-thermal fluctuations averaged in space after the system has reached a steady-state. The top panel shows the distributions for $l_z = 4.74$ $mm$ and the bottom panel for $l_z = 5.02$ $mm$. Panels $a)$ and $c)$ plots the scaled-fluctuation frequency counts for the patterned region, $P$ with a kernel density estimate (dashed). Panels $b)$ and $d)$ plot the scaled-fluctuation frequency counts for the non-patterned annular region, $R$ with a normal curve fit centered at zero (solid).}
    \label{fig7}
\end{figure}

Figure~\ref{fig6} presents the time-averaged scaled thermal variation at steady-state over a region of the film. This scaled thermal variation is calculated by the determining the difference between the temperature of a given pixel from $\langle T\rangle$ of the region of interest then scaled by the same mean, $\delta T^\star = \frac{T_{ij} - \langle T\rangle}{\langle T\rangle}$. Once at steady-state, a series of images (a movie) is recorded at $30$ frames/sec for $15$ minutes. A fixed region of interest is then identified on the image, either one near the edge exhibiting no pattern or one over a hot or cool part of a convection cell, and $\delta T^\star$ is then averaged over $27,000$ frames,

\begin{equation}
\delta T^\star = \frac{1}{T}\int_0^T\delta T^\star (t) dt.
\label{eqn2}
\end{equation}

In Figure~\ref{fig6}a and~\ref{fig6}c, the time-averaged distributions for the upward (hot) and downward (cool) plumes denoted by, $P_{hot}$ and $P_{cold}$ respectively, are shown. In Figure~\ref{fig6}b presents the time-averaged distribution for the entire patterned region, $P$. Each of these three histograms are fitted with a normal distribution function centered at zero. The plots in Figure~\ref{fig6} are shown in a semi-logarithmic scale to highlight the behavior in the tails where deviations from the fit would be most apparent. For the histogram statistics on the hot regions in Figure~\ref{fig6}a, the normal curve describes the data very well over the entire range. However, in the cold regions shown in Figure~\ref{fig6}c, the normal curve does not reproduce the data, especially in the tails, as well and would suggest the possible presence of higher moments to the distribution. The combined distribution is dominated by the hot regions and so does not reveal the deviations from normal as well. As the chosen hot and cold regions do not contain the pattern, they are not influenced by the thermal gradients across a cell therefore, the statistics therein measure pure thermal fluctuations; while the distribution over the whole pattern contains both gradients and fluctuations. Normal distributions imply that the fluctuations are essentially random in nature and that this indicates equilibrium-type fluctuations, which supports the notion that the individual hot and cool regions are each equilibrium-like domains but at different mean temperatures that co-exist in steady-state.

\subsection*{Spatial Analysis}

In Figure~\ref{fig7}, the space-averaged scaled-thermal variation density from the steady-state images are plotted. A steady-state image is chosen in which structures are clearly visible. The two regions of interest, the patterned region ($P$) and the annular non-patterned region ($R$) are chosen. A measure, $\mu$ is defined over the collection of pixel-points in $P$ and $R$ such that,

\begin{equation}
\delta T^\star = \frac{1}{\mu(P)}\int_\mu\delta T^\star (P).
\label{eqn3}
\end{equation}

The left panels ($a$ and $c$) in Figure~\ref{fig7}, report the histograms and the kernel density estimates for the patterned region for the two thicknesses. The salient feature of the plots is the presence of a bimodal behavior. For the same sample under same physical conditions, when a non-patterned region is chosen (right panels, $b$ and $d$), the histograms of the fluctuations are well fitted by a Gaussian distribution function. This bimodal result of the patterned region has two important aspects: i) the ergodicity is clearly broken, and ii) the ergodicity is broken \emph{spatially}. It is interesting to note that a similar bimodal distribution of local thermal fluctuations was reported earlier, but in a very different context~\cite{kadanoff2001turbulent, du2000turbulent}. In the convective cell region ($P$), the distribution contains both gradient and fluctuation contributions to the temperature spatial variation while the hot or cool or ring regions (i.e. chosen regions without a pattern) have a normal distribution. Of course, the emergence of these modes can be attributed to the steady-state patterns of convective instabilities arising due to the upward and downward drafts~\cite{cross1993pattern, nicolis1977self, koschmieder1993benard}. As seen in Figure~\ref{fig7}b and~\ref{fig7}d, the peaks in the distribution are equidistant from the origin with a local minima close to the origin. 

\begin{figure}[t]
    \centering
    \includegraphics[scale=0.6]{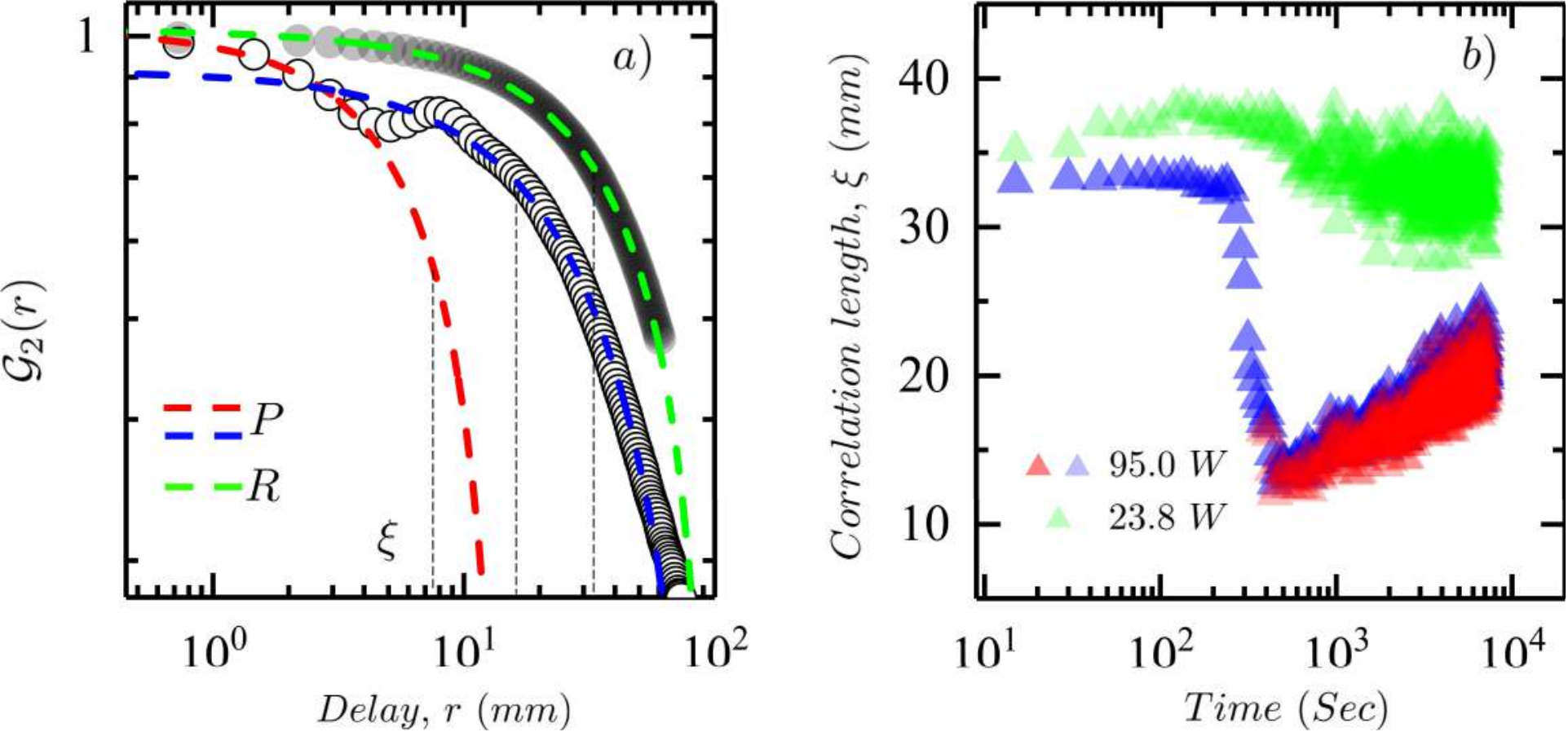}
    \caption{a) Figure shows the two-point autocorrelation function, $\mathcal{G}_2$ as function of distance, $r$ with exponential fits, $\mathcal{G}_2 (r)\sim exp(-\frac{r}{\xi})$ where $\xi$ is the correlation length on a log-log scale. The data shown in grey filled-circles with a single fit is for the non-patterned region of interest ($R$), whereas the data shown in white-filled circles with two fits is for the region of interest that shows emergent structures ($P$). The shown analysis is run on a steady-state image for a $4.74$ $mm$, $95$ $Watt$ sample at steady-state. b) Figure shows the time-dependence of the correlation length for a $4.74$ $mm$ sample at $23.8$ $Watt$ (green triangles) and $95$ $Watt$ (red and blue triangles) as it evolves from room-temperature equilibrium to an out-of-equilibrium steady-state on a semilog scale.}
    \label{fig8}
\end{figure}

In Figure~\ref{fig8}, the pattern of convective cells were characterized spatially by a tracking typical length-scales that emerge in the patterns as the system evolves on heating to an out-of-equilibrium steady-state. Length-scales were extracted from each image using a spatial two-point autocorrelation function, $\mathcal{G}_2$, analysis on the thermal images. The spatial correlation function is defined as, $\mathcal{G}_2 (r) = \langle T(R)\cdot T(R+r)\rangle-\langle T(R)\rangle\langle T(R+r)\rangle$, where $T(R)$ represents the temperature at an arbitrary location on the image, $R$, and $T(R+r)$, the temperature at a distance, $r$ from $R$. A typical two-point autocorrelation function is shown in Figure~\ref{fig8}a for a patterned and un-patterned image, $P$ and $R$. The white filled-circles show the correlation data for the non-patterned region, $R$, described by a single exponential decay fit of the form, $\mathcal{G}_2 (r) = C_1 \exp(-\frac{r}{\xi}) + C_0$. A correlation length ($\xi$) of $33$ $mm$ is estimated from the exponential fit for the $4.75$ $mm$ sample at $95$ $W$ in the non-patterned region, $R$. Whereas, for the patterned region, $P$, two correlation lengths are obtained, $\xi = 18.5$ $mm$ and $9.3$ $mm$. These lengths characterize the average length and width of the observed structures that appear worm-like in nature. Smaller correlation lengths imply increased heterogeneity, the thermal surface of the film becomes progressively structured in time. This is clearly visible from the thermal images shown in Figure~\ref{fig2}b.

\begin{table}[ht]
\centering
\caption{Table shows the calorimetric data from the steady-state images at different powers for the two thickness ($l_z = 4.74$ $mm$ and $5.02$ $mm$). The numbers listed in the first column denote the specified points in the plots shown in Figure~\ref{fig9}. The top temperature ($T_{top}$) is recorded by the thermal camera, bottom temperature ($T_{bottom}$) by the thermocouple $T_2$, the hot and cold spot temperatures ($T_{P_{hot}}$ and $T_{P_{cold}}$) are obtained by spatially averaging regions of interest ($P_{hot}$ and $P_{cold}$) from the thermal images, conduction temperature ($T_{cond}$) is calculated from Equation~\ref{eqn4}, and the Rayleigh Number ($Ra = \frac{g\beta l_z^3}{\nu\alpha}(T_{bottom} - T_{top})$) from the listed values in Table~\ref{table1}.}
\label{table2}
\begin{tabular}{|l|l|l|l|l|l|l|l|l|}
\hline
$l_z$ & $\#$ & \emph{Power} & $T_{top}$ & $T_{P_{hot}}$ & $T_{P_{cold}}$  & $T_{bottom}$ & $T_{cond}$ & \emph{Rayleigh Number}\\
$(mm)$ & &  $(W)$ & $(^\circ C)$ & $(^\circ C)$ & $(^\circ C)$ & $(^\circ C)$ & $(^\circ C)$ & $Ra$\\
\hline
& $1$ & $23.8$ & $39.4$ & $--$ & $--$ & $53.2$ & $46.8$ & $831$ \\
& $2$ & $42.2$ & $48.4$ & $61.5$ & $54.8$ & $71.7$ & $61.7$ & $1410$ \\
4.74 & $3$ & $66$ & $59.9$ & $78.2$ & $69.7$ & $89.5$ & $76.1$ & $1790$ \\
& $4$ & $95$ & $70.9$ & $100.9$ & $91.1$ & $115$ & $96.4$ & $2670$ \\
& $5$ & $130$ & $89.8$ & $124.8$ & $114.1$ & $147$ & $122.2$ & $3464$ \\
\hline
& $1$ & $10.5$ & $30.3$ & $--$ & $--$ & $37.9$ & $34.5$ & $535$ \\
& $2$ & $23.8$ & $38.1$ & $43.1$ & $39.7$ & $53.4$ & $46.9$ & $1080$ \\
5.02 & $3$ & $42.2$ & $47.2$ & $63.5$ & $56.7$ & $70.9$ & $60.9$ & $1670$ \\
& $4$ & $66$ & $58.8$ & $84.4$ & $73.6$ & $91.8$ & $77.7$ & $2330$ \\
& $5$ & $95$ & $73.1$ & $101.3$ & $90.1$ & $115$ & $96.4$ & $2960$ \\
\hline
\end{tabular}
\end{table}

The time evolution of the extracted correlation lengths are shown in Figure~\ref{fig8}b as the system reaches steady-state. At a heating power of $95$ $W$, sufficient for structures to emerge, a single correlation length is seen initially as long as the width of the film as no pattern has emerged just as was seen over the entire time-evolution of the film heated at $23.8$ $W$ that never exhibited emergent structures. The single correlation length remains roughly constant ($\xi\sim 30 - 35$ $mm$) until about $200-300$ seconds, when it suddenly decreases to about $10-12$ $mm$ with the emergence of a second length-scale $\sim 9-10$ $mm$ (shown in solid red triangles). Note that the sharp drop in the correlation length as a function of time coincides with the drop in standard deviation of the temperature plots during the heating process (see Figure~\ref{fig4}a). The data in solid green triangles is for the film heated by $23.8$ $W$, and illustrates the result of this spatial analysis as a reference since this run exhibits no patterns. However, this analysis is limited, as seen in both Figures~\ref{fig8}a and~\ref{fig8}b, by the inherent noise of the correlation data due to limited spatial span available and the limited thermal/spatial resolution of the camera. Thus, these results are only estimates of the true correlation statistics.

\begin{figure}[t]
    \centering
    \includegraphics[scale=0.6]{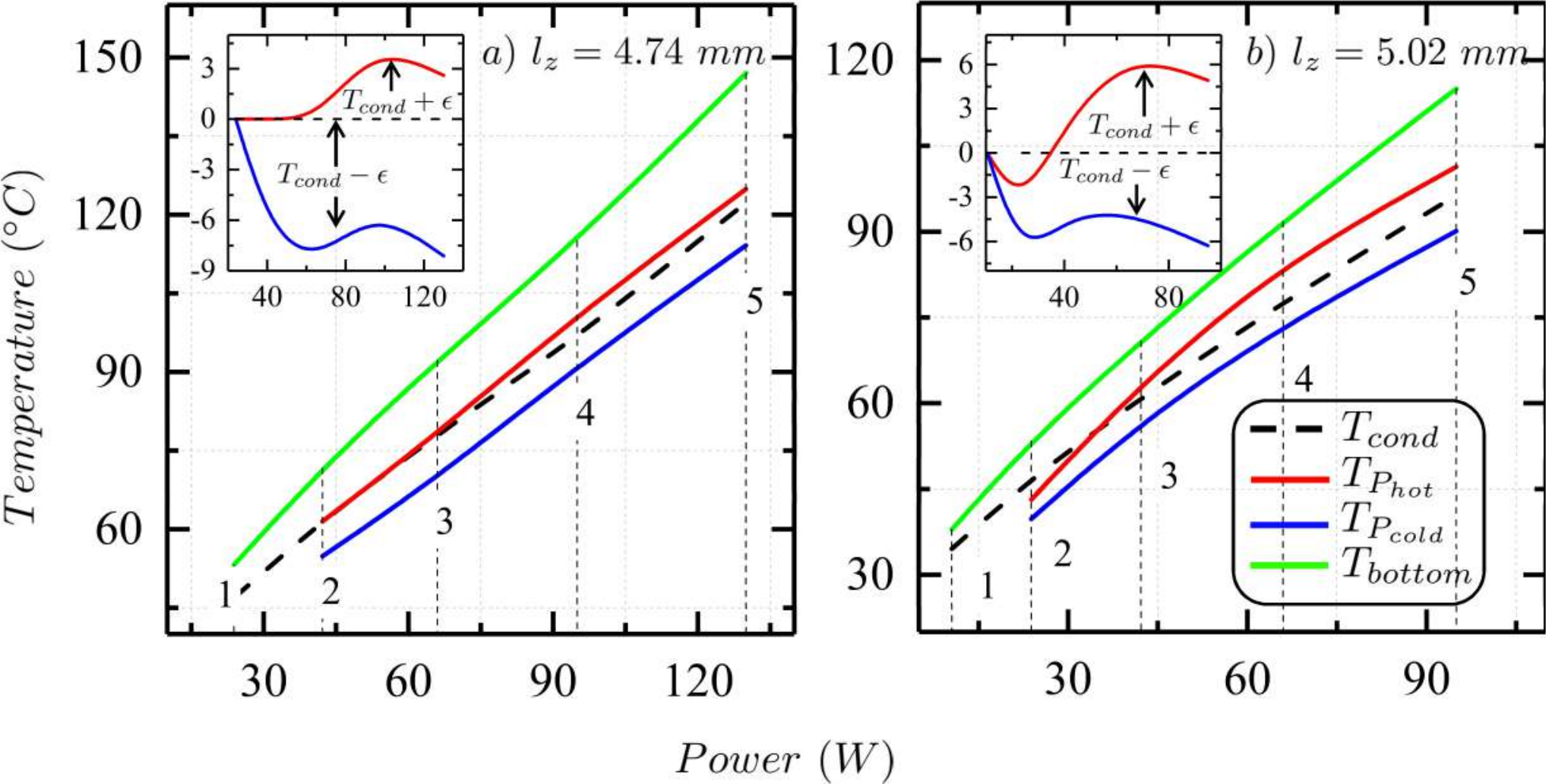}
    \caption{Figure shows the temperature plots ($T_{P_{hot}}, T_{P_{cold}}, T_{cond}$ and $T_{bottom}$) for the steady-state images at different values of input power for a) $l_z = 4.74$ $mm$ and b) $l_z = 5.02$ $mm$. The inset plots capture the variation in the plume temperatures ($T_{P_{hot}}$ and $T_{P_{cold}}$) about the theoretical conduction temperature ($T_{cond}$) as a function of power. For details about the specific points denoted in the plots, refer Table~\ref{table2}. Also, note that $\epsilon$ is arbitrary.}
    \label{fig9}
\end{figure}

Finally, the experimental configuration allows the comparison of the temperatures across the film surface to that expected if convection was absent. That is, to what would have been the theoretical temperature of the top surface of the fluid film if the mechanism of heat transport had been through pure conduction. In order to calculate the theoretical conductive temperature, $T_{cond}$, the steady-state heat conduction equation is used along with the available calorimetry data,

\begin{equation}
\dot{Q} = \frac{(m_{Cu}c_{p_{Cu}} + m_{oil}c_{p_{oil}})(T_{bottom} - T_{top})}{2\times 60\times 60} = -kA\nabla T = -kA\Big(\frac{T_{cond} - T_{bottom}}{l_z}\Big),
\label{eqn4}
\end{equation}

where $A$ is the area of the copper pan, the material properties are given in Table~\ref{table1}, and the measured temperature values from Table~\ref{table2} of. the theoretical expected temperature ($T_{cond}$) and the temperature of the upward and downward drafts ($T_{P_{hot}}$ and $T_{P_{cold}}$). The resulting values of all the temperatures are listed for both the thicknesses in Table~\ref{table2} and plotted as a function of applied power in Figure~\ref{fig9}. The critical Rayleigh Number for structures to emerge is $1708$ and for experiments beyond this critical value (see Table~\ref{table2} last column, after third row), the theoretical conduction temperature is close to the weighted average of the hot and cold plume temperatures denoted by, $T_{P_{hot}}$ and $T_{P_{cold}}$. In the inset of Figure~\ref{fig9}, the variance of the plume temperatures about the conduction temperature ($T_{cond} + \epsilon$ and $T_{cond} - \epsilon$) as a function of the applied power is shown. Interestingly, the nature of this variation does not follow a linear relationship, but rather oscillates above and below $T_{cond}$ almost anti-symmetrically. Although macroscopically the system is at steady-state, and so time-invariant, the spatial regions corresponding to $T_{P_{hot}}$ and $T_{P_{cold}}$, can be thought of as separate local equilibrium-like regions coexisting with each other.

\section*{Discussion}

\begin{figure}[t]
    \centering
    \includegraphics[scale=0.6]{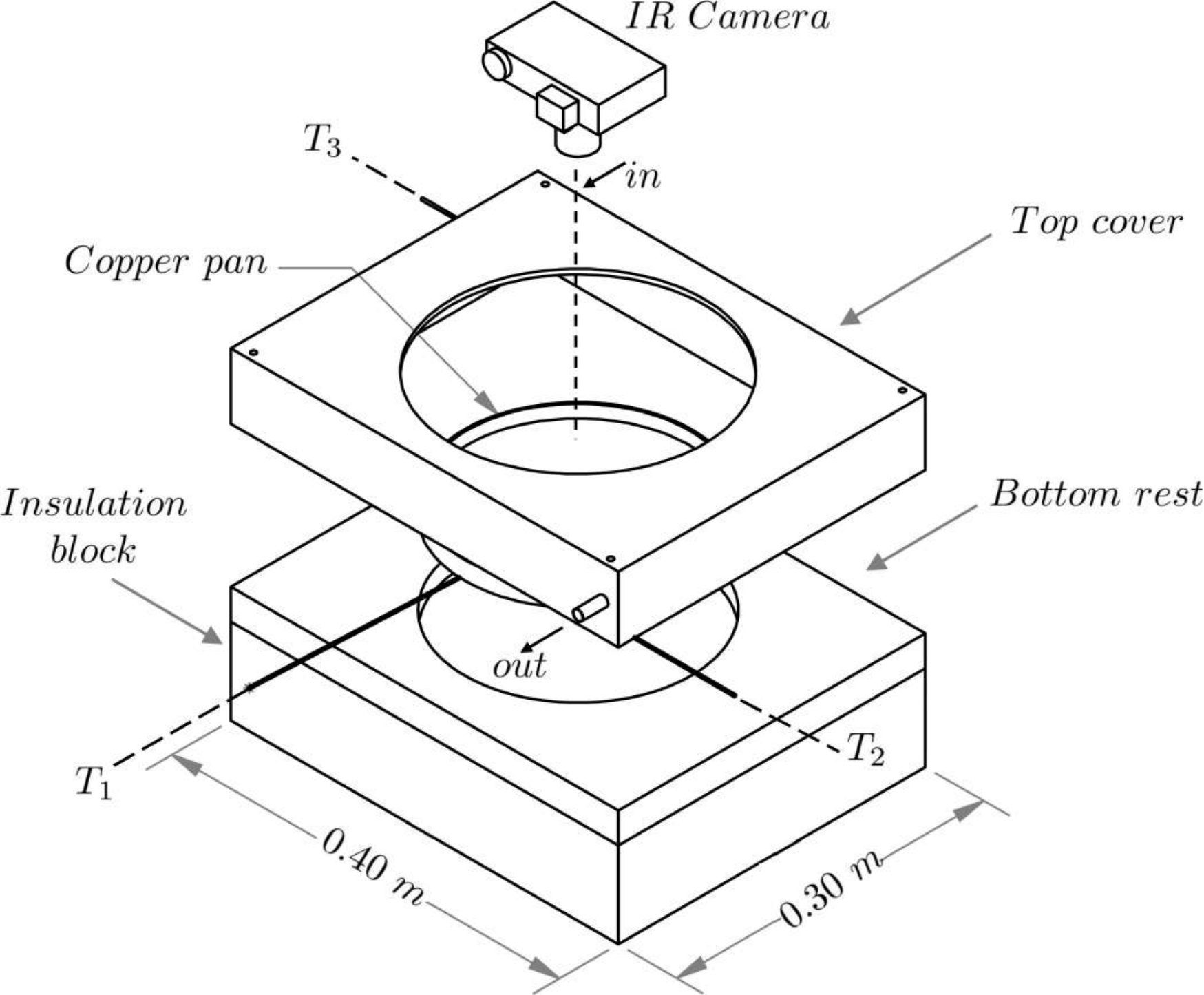}
    \caption{Figure illustrates the experimental setup with the copper pan ($2R = 0.225~m$), the three thermocouples ($T_1$, $T_2$, $T_3$), inlet and outlet ducts for the forced convective heat transfer, and the Infra Red camera for real-time thermal imaging. The inlet and the outlet ducts are present on the top cover and the copper pan sits on a wooden bottom rest and a polyurethane foam foundation which acts as an insulator.}
    \label{fig1}
\end{figure}

The lack of a theoretical framework makes systems that are out-of-equilibrium difficult to study. However, the Rayleigh-B{\'e}nard convection, with controllable system variables and access to all measurable quantities is an attractive platform to shed light that may guide theoretical development. In this study, the Rayleigh-B{\'e}nard system is used as a prototype to gain insights about far-from-equilibrium thermodynamics. Equilibrium behavior is typically easy to visualize, as at equilibrium, all macroscopic thermodynamic variables collapse into fixed points in phase-space~\cite{planck2013treatise, clausius1854veranderte, gibbs1906scientific}. Temperature, which plays a key role in equilibrium thermodynamics, is often quoted as a bad thermodynamic variable to characterize far-from-equilibrium systems, and hence should not be used to describe out-of-equilibrium behavior. This notion is technically sound, as macroscopic variables when far-from-equilibrium are constantly changing in time and no descriptive state-function can be written. Although, when deviations are linear and relatively small, the equilibrium description can be extended under the claims of local equilibrium hypothesis. Nevertheless, even after $200$ years of effort, a general theory of far-from-equilibrium thermodynamics is currently missing, and is still quoted as ``work in  progress"~\cite{vilar2001thermodynamics, kolmogorov1941local, wikipedia}. The argument against the use of temperature as a measure to theorize far-from-equilibrium thermodynamics although logically valid does not provide a way to solve this long-standing problem. This work seeks to provide experimental observations to stimulate theoretical progress.

A remarkable observation from our analysis of the steady-state thermal images is that local equilibrium-like regions appear to spatially coexist in an out-of-equilibrium system driven presumably by the partitioning of the heat energy flow into entropic and coherent work (the convection circulation). The system is therefore non-ergodic as a whole, but is ergodic in equilibrium-like sub-regions that do not exhibit a pattern in time, but not over the entire film. Since, time translation symmetry is preserved, any macroscopic description of the system should be found to conserve energy (or have applicable the First Law of Thermodynamics). As translation symmetry is broken over the whole film, there must exist internal gradients of temperature between adjacent regions, the internal coherent work that drives the convective flow of fluid is also maintaining these internal temperature gradients. The Second Law is well preserved for the macroscopic description of the system, locally however it gets violated due to the emergence of structures and internal gradients~\cite{lieb1998guide, martyushev2006maximum, onsager1931reciprocal, tolman1948irreversible, wehrl1978general, sharma2007natural, kaila2008natural}. This can be seen in the cooling profiles in Figure~\ref{fig5}, where the structures and internal gradients disappear as soon as the system relaxes back to room temperature. Future work on this system would naturally be to quantitatively determine the amount of work required to maintain these co-existing localized gradients on the energy manifold. The emergent work averaged over these states and the free-energy differences between the equilibrium-like states are statistically related as, $\exp(-\langle W\rangle/k_B T) = \exp(-\Delta F/k_B T)$~\cite{jarzynski1997nonequilibrium, jarzynski1997equilibrium}. Insights about the free-energy of the local equilibrium-like states would throw considerable light on the interpretation of the partition function for such non-equilibrium steady-state systems~\cite{yamada1967nonlinear, Gallavotti2016, gallavotti2019nonequilibrium}.

The breaking of translation symmetry and unbalanced internal gradients can possibly explain the peculiar nature of the standard deviation plots during heating and cooling. From Figure~\ref{fig4} and Figure~\ref{fig2} we can observe that the curve for the $T_top$ bends at time, $t \sim 200-300$~sec when the first structures appear. It is tempting to conclude that the two observations are related to the same phenomenon. As the system is heated, local equilibrium-like regions start to emerge which causes the system to start getting correlated. As the correlations get stronger, the system starts behaving as collections of local equilibrium-like domains (see Figure~\ref{fig8}). As the fluctuations between these domains get stronger (compare the ranges of the scaled fluctuations from the $x-$axis in Figure~\ref{fig7}), they start dominating the fluctuations elsewhere which gives rise to Casimir like effect~\cite{vella2017fluctuation, chatterjee2018non, yadati2018detailed, chatterjeeccs17}. Due to the finite size of the system these effects propagate at a much faster rate than mere thermal diffusion. This is readily observed in the sudden decline of the standard deviation during heating. While cooling, the domains disintegrate and the system becomes weakly correlated, thus the strong fluctuations almost immediately disappear.

In conclusion, although macroscopically the system is at steady-state the regions in space corresponding to $P_{hot}$ and $P_{cold}$ can be realized as localized heat baths with equilibrium-like statistics confined within them. The dissipation from the equilibrium fluctuations within these localized regions manifests as a spatial variation of the temperature manifold, the curvature of which indicates how far one is from the equilibrium state, $T_{cond}$ (see Figure~\ref{fig9}). The upward and downward drafts at these localized regions perform internal work to maintain the convection (structure and internal gradients) while resisting spontaneous equilibration. An intuitive understanding of this mechanism is the bifurcation of the theoretical conduction temperature beyond the critical Rayleigh Number (see Figure~\ref{fig9}). Therefore, in order to interpret temperature far-from-equilibrium we must consider, temperature not as state variable but as a functional on the energy landscape~\cite{chatterjee2018many}. This energy landscape consists of local equilibrium-like points, and within each of these regions the macroscopic equilibrium thermodynamics ideally holds true. A theory that would encompass this idea must have to preserve the First Law while modifying it to include the emergence of internal gradients~\cite{chatterjee2016thermodynamics, garcia2008thermodynamics, chatterjee2013principle, srinivasarao2019biologically, karl2013tuning}. The results presented in this paper may provide a new perspective and a way forward to laying out the foundations for a theoretical interpretation of far-from-equilibrium phenomena.

\section*{Methods}

\subsection*{Experimental Methodology}

A thin layer of silicone oil is heated in a copper pan whose average diameter is $0.225$ $m$. The thermal and material properties of the oil is outlined in Table~\ref{table1}. The pan is heated from the bottom by an electric heater. In Figure~\ref{fig1} we illustrate the experimental setup in detail. The top cover is made up of wood and has inlet and outlet ducts for forced convective heat transfer. The two thermocouples $T_2$ and $T_3$ measure the temperature of the incoming and outgoing gas respectively. The bottom rest, also made up of wood has a cavity with a recess on which the copper pan sits snugly. The wooden base rests on top of a block of Polyurethane foam. The thermocouple, $T_1$ is connected to the base of the copper pan which measures the bottom temperature of the pan ($T_{bottom}$). An infra-red camera (with a precision $\sim 10^{-3}K$), placed concentrically above the copper pan captures the real-time thermal images from a height. The temperature scale of the camera is calibrated by heating the empty copper pan. Due to small varying thickness of the base of the copper pan, the film thickness and the surface temperature of the top is averaged over the entire exposed area. The system is heated by regulating the power input through the heater. The resistance of the electric heater is $37.5\pm 0.5$ $\Omega$. At a specific power, the system is let to evolve over time such that the mean bulk-temperature stops fluctuating. Once the system reaches a steady-state (after approximately two hours), the mean temperature of the top surface is denoted by $T_{top}$.

\begin{figure}[hb!]
\centering
\includegraphics[scale=0.5]{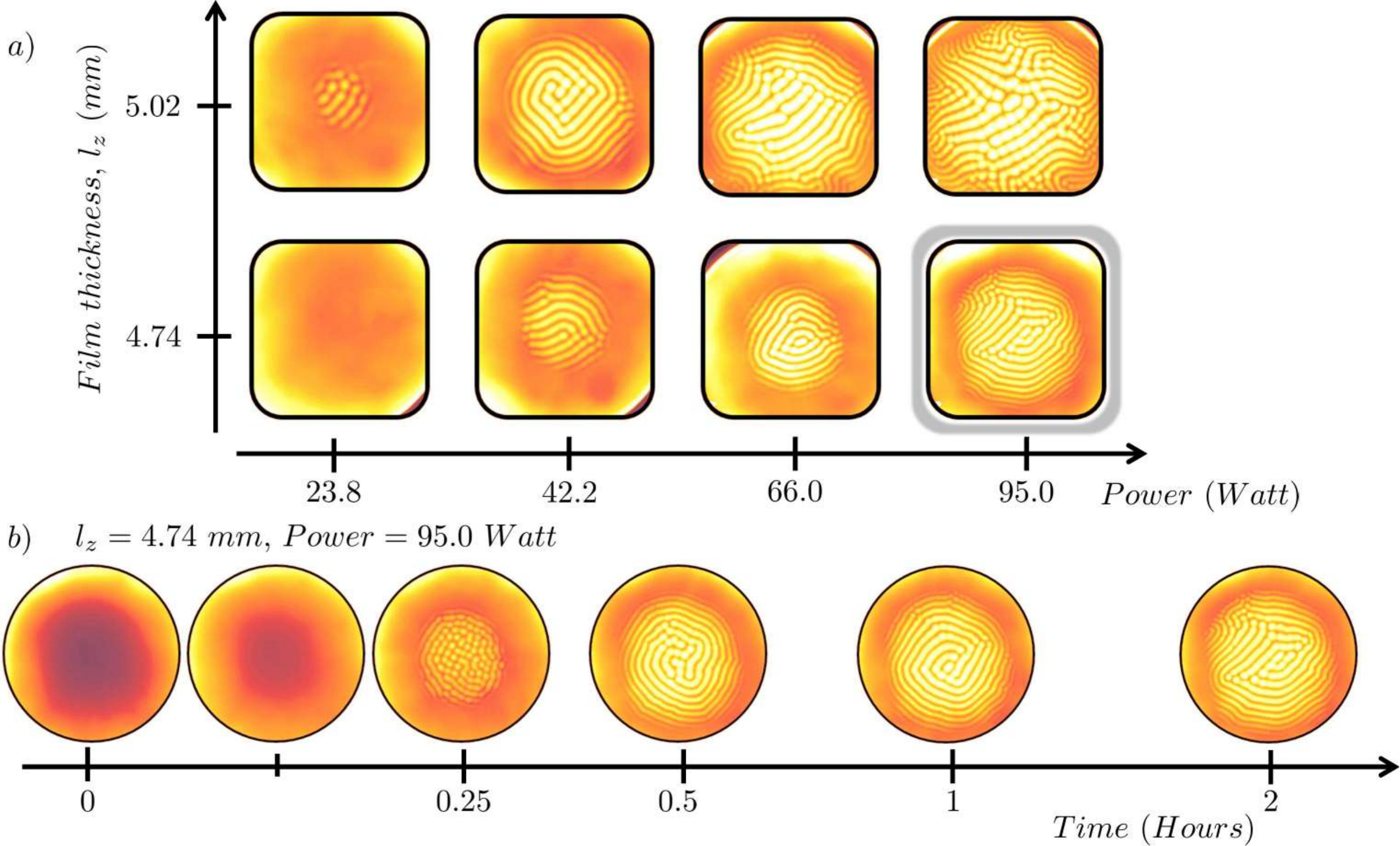}
\caption{a) Figure shows steady-state thermal images recorded for two thickness, $l_z = 4.74$ $mm$ and $5.02$ $mm$ at various powers. b) Figure shows the time-evolution of the $l_z = 4.74$ $mm$ at $95.0$ $W$ over a period of two hours. Note that the shown images are logarithmically placed in time.}
\label{fig2}
\end{figure}

It is important to note that the apparatus was not intended for high control of convection cells. Rather, the goal was to have convection cells over as wide as an area possible for the thermal imaging to yield significant temperature statistics, both temporally and spatially. The criteria was then for a large diameter to thickness ratio of the apparatus ($225$ $mm$ / $7$ $mm$) that yielded a stable convection cell pattern at least or greater than $150$ $mm$ in diameter and stable for as long as the power is applied~\cite{pandey2018turbulent, scheel2017predicting, bodenschatz2000recent}. To the best of our knowledge, this has not been done before. Therefore, the thrust is to shed light onto the relationship between the emergence of structure in driven out-of-equilibrium systems and the far-from-equilibrium definition of temperature.

\begin{table}[hb]
\centering
\caption{Table outlines thermal and material properties of the Silicone oil sample that was used to perform the current study~\cite{shinetsu}.}
\label{table1}
\begin{tabular}{|l|l|l|l|l|l|}
\hline
\emph{Kinematic Viscosity}  & \emph{Density}  & \emph{Thermal Conductivity} & \emph{Specific Heat} & \emph{Thermal Diffusivity} & \emph{Compressibility}\\
$\nu$ $(cSt)$ & $\rho$ $(Kg/m^3)$ & $k$ $(W/m-K)$ & $c_{p_{oil}}$ $(J/Kg- K)$ & $\alpha$ $(m^{2}/s)$ & $\beta_T$ $(m^{2}/N)$\\
\hline
$150$ & $970$ & $0.16$ & $1500$ & $1.099\times 10^{-7}$ & $9.5\times 10^{-4}$ \\ \hline
\end{tabular}
\end{table}

In Figure~\ref{fig2}a we plot the steady-state thermal images of the convection patterns for two film thickness ($l_z = 4.74$ $mm$ and $5.02$ $mm$) with increasing power, along $x-$axis. Each image has a color scheme that is a function of its independent thermal scale (recorded by the calibrated Infra Red camera). As every steady-state image is embedded with its own calorimetric information, the temperature at each pixel location can be computed through a simple linear interpolation that transforms the grey-scale bit value to a corresponding temperature. In Figure~\ref{fig2}b we present a graphical representation of the time-evolution of the patterns for the film thickness, $l_z = 4.74$ $mm$ at $95.0$ $W$ (highlighted in Figure~\ref{fig2}a). The images are placed logarithmically along the $x-$axis to bring out the clear difference in time taken by the system before and after the onset of patterns. In the pre-pattern (or no pattern) stage, the dynamics of the system is very fast, specially during the first quarter of the hour. However, once structures start emerging the dynamics of the system slows down drastically, and during the last hour it barely shows any measurable dynamical changes as is clearly visible in Figure~\ref{fig2}b.

\subsection*{Analysis}

\begin{figure}[t]
    \centering
    \includegraphics[scale=0.6]{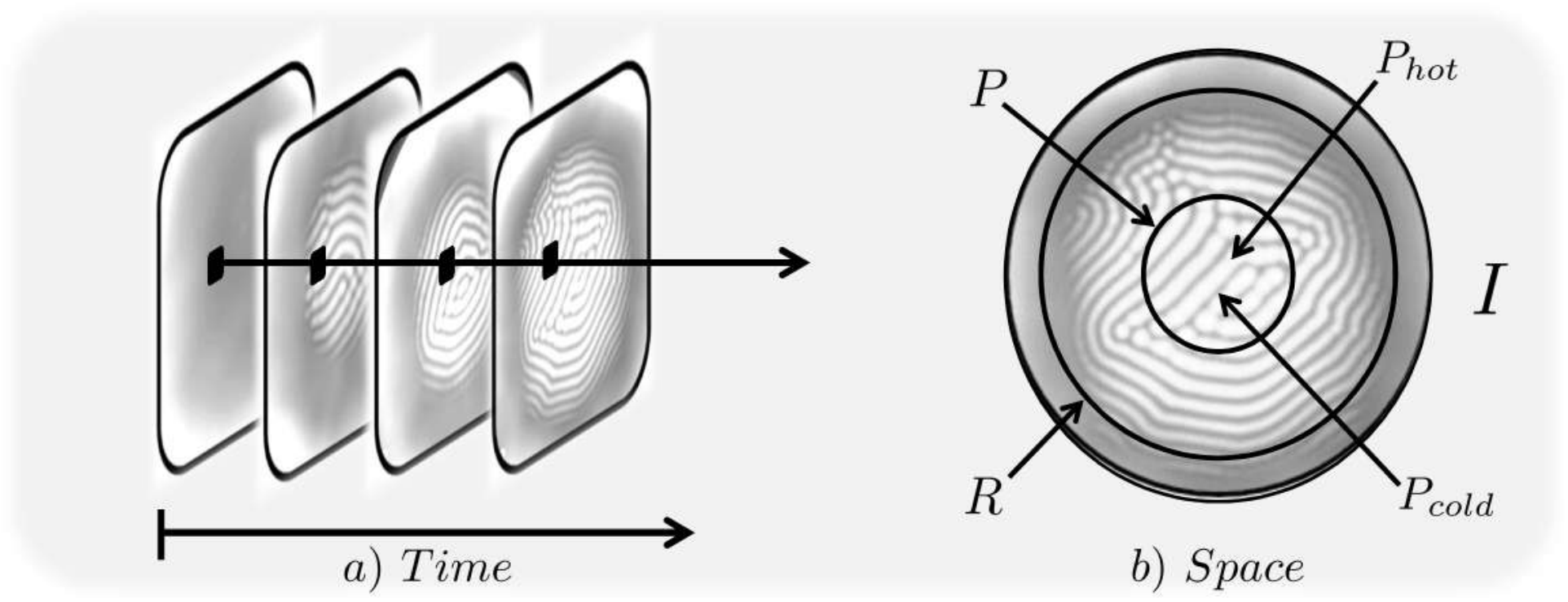}
    \caption{a) Figure illustrates the temporal analysis of an arbitrary region of interest on the images as a function time as the system evolves from room temperature equilibrium to an out-of-equilibrium steady-state. b) Figure shows the regions of interest for the spatial analysis on the steady-state image of a Rayleigh-B{\'e}nard convection. The complete image is denoted by $I$, the annular region without any structures by $R$, the circle at the center by $P$, the upward (bright spots) and downward plumes (dark spots) by $P_{hot}$ and $P_{cold}$ respectively.}
    \label{fig3}
\end{figure}

A sample of the raw images that were recorded by the Infra Red camera are shown in Figure~\ref{fig2}a and~\ref{fig2}b. These raw images ($I$) are then converted into a $N\times N$ matrix of temperature, where each entry of the matrix element ($I_{ij}$) corresponds to the temperature of each pixel ($T_{ij}$) on the image. These images are then statistically analyzed both spatially and temporally. In Figure~\ref{fig3} we depict the two types of analysis that are performed on these images. In Figure~\ref{fig3}a we perform a temporal analysis of the images as the system evolves to a steady state. An arbitrary region of interest is identified and is then followed in time. The statistics that are obtained, are then analyzed a function of time or are averaged over time. In Figure~\ref{fig3}b, we spatially analyze the steady-state images as obtained from the thermal camera. The analysis of this type gives us insights about the spatial aspects of the system once steady-state has been achieved and structures have emerged. The two primary regions of interest in this type of analysis are the patterned region ($P$) and the non-patterned region (or the ring region) $R$. Within the patterned region, $P$, the brighter spots represent upward plumes and are denoted by $P_{hot}$, while the darker spots represent downward plumes, and are denoted by $P_{cold}$.

\section*{Acknowledgements}
The authors are indebted to their collaborator Georgi Y. Georgiev (Assumption College) and the contributions of Sean McGrath. The authors are thankful for the support of the Department of Physics at Worcester Polytechnic Institute. Finally, the authors are extremely grateful to the excellent comments and criticisms from the editor and the reviewers.

\section*{Author contributions statement}

A.C. and G.I. conceived the experiment, designed the methodology, and wrote the manuscript, A.C. and Y.Y. conducted the experiments, A.C., N.M. and Y.Y. analyzed the results, Y.Y. prepared Figure 8 and A.C. prepared the remaining figures to final form, G.I. approved the manuscript, supervised and funded the research. All authors reviewed the manuscript.

\end{document}